\documentstyle[psfig,prl,aps]{revtex}

\def\be{\begin{equation}}
\def\ee{\end{equation}}
\def\ba{\begin{array}}
\def\ea{\end{array}}
\def\<{\langle}
\def\>{\rangle}
\def\~{\tilde}

\def\firugh {\nabla\cdot\left({\nabla H\over \parallel\nabla H\parallel ^2}
             \right)}
\def\psirugh {\nabla\cdot\left[\nabla\cdot\left({\nabla H\over \parallel
              \nabla H\parallel ^2}\right){\nabla H\over \parallel
              \nabla H\parallel ^2}\right]}             
\def\dekinchin {{d\sigma\over \parallel\nabla H\parallel}}
\def\gradh {{\nabla  H\over \|\nabla H\| }}
\def\pquadro {\sum_{i=1}^{N} p_i^2\over N}
\def\derome {{\partial\omega\over \partial E}} 
\def\derrome {\partial^2\omega\over\partial E^2}
 1
\begin{document}
\title{Ergodic Properties of Microcanonical Observables}
\maketitle
\date{\today}
\begin{center}
{\large Cristian Giardin\`a}\\

{\small\it
Dipartimento di Fisica Universit\`a di Bologna and INFN Sez. di Bologna,\\
via Irnerio 46, 40126 Bologna, Italy\\
{\rm giardina@bo.infn.it }\\
}

\vspace{0.3cm}
{\large Roberto Livi }\\
{\small\it
Dipartimento di Fisica dell'Universit\`a, INFN Sez. di Firenze and INFM
Unit\`a di Firenze,\\
Largo E. Fermi 2, 50125 Firenze, Italy\\
{\rm livi@fi.infn.it}
}
\end{center}

\vspace{0.3cm}
\begin{abstract}
\noindent {\bf Abstract:}
The problem of the existence of a Strong Stochasticity Threshold in the 
FPU-$\beta$ model is reconsidered, using suitable microcanonical observables 
of thermodynamic nature, like the temperature and the specific heat. 
Explicit  expressions for these observables are
obtained by exploiting rigorous me\-thods of differential geometry.
Measurements of the corresponding temporal 
autocorrelation functions locate the threshold at a finite value
of the energy density, that results to be independent of the number
of degrees of freedom.

\vspace{0.2cm}
\end{abstract}
\section{Introduction}
Models of interest for applications in plasma and 
condensed matter physics as well as in molecular biology and
chemistry are described by many degrees of freedom Hamiltonians 
of the form
\begin{equation} 
H = \sum_{i=1}^N {{p_i^2}\over{2m}} + V(\{q_i\})
\label{hamilton}
\end{equation}
where $V(\{q_i\})$ is a nonlinear interaction potential among
the $N$ particles of mass $m$, located on a regular lattice,
whose sites are labelled by the integer index $i$.
The displacement and momentum canonical coordinates of the
$i$-th particle are denoted by $q_i$ and $p_i$, respectively. 
Apart ramarkable exceptions like the Toda chain 
\cite{toda}, 
a generic choice for $V$ yields a chaotic dynamics.
One of the main issues in this field is the existence
of different levels of chaos that typically occur when some
control parameter (e.g. energy or energy density) is varied.
Fermi, Pasta and Ulam in their pioneeering numerical experiment 
\cite{FPU}, 
first observed the strong rigidity of low-$k$ 
excitations that prevented equipartion of the energy among the 
Fourier modes over exceedingly large time scales.  
Izrailev and Chirikov  provided an explanation of this fact in terms of 
the resonance-overlap criterion
\cite{IC}, 
while more refined  numerical experiments
showed that, for sufficiently high energies, equipartition among
the Fourier modes sets in very rapidly
\cite{CIT,BSBL,CCDGS,BT}.
In particular, already  Bocchieri et al.
\cite{BSBL} 
raised the question of the existence of an energy threshold separating
a quasi-regular dynamics from a highly chaotic phase. Nowadays, there
is theoretical and numerical evidence that such a threshold does exist
for large values of $N$. In this respect, it is worth stressing the
following points:
\begin{itemize}
\item this threshold in general occurs at energy values that
prevent any possibility of quantitative explanation in terms of the
canonical perturbation theory;
\item  in the  strongly chaotic phase the time needed to
reach equipartition is almost independent of energy and $N$;
just beyond the threshold it suddenly increases, while the dynamics results 
to be very weakly chaotic.  \end{itemize} 
These remarks point out that the equipartition threshold (ET) separates 
definitely different dynamical regimes, in such a way that
for sufficiently small energies the time needed for approaching
equilibrium properties may become so long that the dynamics maintains
an ordered structure over any practically available time scale.
\cite{CLMP}.
On the other hand, it should be stressed that ET does not seem to 
exhibit analogies with the standard scenario of equilibrium
phase transitions: for instance, it is not at all 
clear which kind of symmetry
breaking mechanism, if any,  could be responsible for the slowing down of
energy equipartition among the Fourier modes. 

A further crucial problem
concerns the persistence of ET in the thermodynamic limit.
This is a key point that recently led various authors to 
reconsider this problem, by exploiting different approaches and techniques.
Let us briefly summarize the state of the art.
Some investigations have refined the original approach based on the study 
of energy equipartition among the Fourier modes:  the so-called 
spectral entropy was first introduced as a global
observable able to quantify the amount of energy equipartition
\cite{LPRSV,LPRV}.
The scaling properties of this quantity have been carefully analyzed
in ref.s \cite{KLR,DLR}, yielding the conclusion that 
the time needed to reach energy equipartition 
scales in a nontrivial way with both the energy and $N$.
Further numerical investigations and theoretical arguments
based on an improved treatment of the Chrikov's resonance overlap
criterion 
\cite{DLR,DLL,She}
indicate that, at least for low-$k$ modes excitations,
ET should vanish in the thermodynamic limit.

An alternative approach to the characterization
of the different dynamical regimes occurring in models
like (\ref{hamilton}) has been pursued by directly
investigating proper geometrical features of 
the phase space. Concepts and tools taken from  Riemannian geometry
were translated in Hamiltonian language in
\cite{P93,CP93,CP95}.
By exploiting such techniques it has been possible to obtain an 
analytic estimate in the thermodynamic limit of the maximum 
Lyapunov exponent as a function of the energy density 
for the FPU $\beta$-model \cite{CLP}.
It exhibits a crossover between a weak and a strong chaotic
regime when the energy density is increased beyond a threshold value,
that has been identified as the Strong Stochasticity Threshold (SST)
\cite{PCS}.
Similar conclusions in favour of the existence of SST in the
thermodynamic limit have been drawn by considering other specific
geometrical indicators, like generalized curvatures of the phase space,
whose reliability has been tested for various models
\cite{ACM93,ACM95,ABCM}.
By the way, theese studies also indicate that in the strongly chaotic
regime these models may still be far from ergodic.

Upon all of these remarks, one might wonder if the existence
of a threshold (either ET or SST)  may have some influence on 
the equilibrium statistical
properties of many degrees of freedom hamiltonian models.
Beyond the evident interest of such a question for the
foundations of Statistical Mechanics, this is a crucial
point for molecular dynamics applications, that usually
assume the validity of the ergodic hypothesis, i.e. the
equivalence of time and ensemble averages.
Such a problem would be solved from the very beginning if one
could prove ergodicity for a model like (1).
Unfortunately, this is possible only in a few special cases.
On the other hand, it is commonly accepted that chaoticity 
should be sufficient for reproducing reliable statistical
predictions for most of the observables of physical interest.
A partial disproval of this assumption was obtained in 
\cite{JST},
where it has been shown that the equivalence between
time and ensemble averages for canonical thermodynamic
observables (e.g. internal energy and specific heat)
mainly depend on the model and on the nature of the observable,
rather than on the degree of chaoticity of the dynamics.

In this paper we aim to clarify most of these open points 
by tackling the problem through the direct comparison of
statistical and dynamical properties of microcanonical
thermodynamic observables. The general formalism yielding
explicit dynamical expressions for such observables
is introduced in Section II, where we also comment on their 
geometrical nature. The results obtained by molecular 
dynamics calculations for the FPU $\beta$-model
are reported in  Section III.
We show that SST is predicted by the decay rate of
the temporal autocorrelation function associated with
the microcanonical temperature and specific heat,
while ergodic-like properties are found to depend mainly
on the observable at hand, irrespectively of the degree
of chaoticity of the dynamics. 
Conclusions and perspectives are contained in Section IV.

\section{Microcanonical Thermodynamic Quantities}

The equivalence in the thermodynamic limit of the predictions obtained from 
different statistical ensembles ($\mu$-canonical, canonical and $G$-canonical)
is a widely accepted and partially proved fact
\cite{Ruelle}.
Explicit expressions for thermodynamic observables
are usually computed by canonical averages. Their relation
with the corresponding microcanonical averages was established in
a celebrated paper by Lebowitz, Percus and Verlet
\cite{LPV}.
Only recently H.H. Rugh, assuming ergodicity of the phase space, 
obtained a general
rigorous expression for the temperature of the microcanonical ensemble,
that results to be strictly related to the 
geometrical structure of the phase space 
\cite{Rugh1}.
Analogously, an explicit formula 
for the specific heat at constant volume 
has been also obtained \cite{Rugh2}.

In what follows we are going to reproduce the expressions
of these microcanonical thermodynamic quantities, by exploiting
a different approach, based on
standard tools of differential geometry (see the Appendix).
We find that any of these quantities is an explicit, even if
in some cases complicate, function of the canonical coordinates.
As a consequence, this geometrical approach provides
a natural bridge between a dynamical and a
thermodynamical description of the microcanonical ensemble.
Upon the remarks of the previous Section, we are interested
in understanding what can be inferred about thermodynamics by
molecular dynamics experiments performed on hamiltonian
models of the form (1). As we have already stressed, in general
such models are known not to be ergodic in a strictly mathematical 
sense. In this perspective, it could be hard to find out more appropriate 
probes for testing ergodicity, than such microcanonical
thermodynamic observables.

In analogy with the rigorous approach of Rugh, we assume 
ergodicity in order to guarantee that the microcanonical
entropy $S$ is a well defined quantity associated to
some hypervolume in phase space, equipped with a uniform
undecomposable probability measure. In the 
${\mu}$-canonical ensemble 
$S$ plays the role of a generalized thermodynamic potential
from which any thermodynamic quantity can be obtained by derivations
w.r.t. the physical parameters energy $E$ and volume $V$:
\begin{equation}
{1\over T}=\left({\partial S\over \partial E}\right)_V 
\quad\quad\quad\quad
{1\over C_V}=\left({\partial T\over\partial E}\right)_V 
\end{equation}

\begin{equation} 
{P\over T}=\left({\partial S\over\partial V}\right)_E 
\quad\quad\quad\quad
{1\over V\alpha_E}=\left({\partial T\over\partial V}\right)_E  
\end{equation}
where $\alpha_E$ is defined as the thermal expansion coefficient
in the $\mu$-ensemble.
In order to be correctly defined $S_{\mu}$ must fulfill the following
properties: 
\begin{itemize}
\item  additivity;
\item  invariance under adiabatic reversible transformations;
\item  it must be a non-decreasing function for irreversible adiabatic
transformations.
\end{itemize}
Two choices are possible for the $\mu$-ensemble:
\begin{eqnarray*}
\mbox{A)} \quad S(E, N, V) &=& \ln (c_N\omega (E, N, V)) =
\ln \left(c_N \int dq\;dp \;\delta (E - H(q ,p))\right) \\
&=& \ln \left(c_N{\int_{H=E}{\dekinchin}}\right)
\end{eqnarray*}
\begin{eqnarray*}
\mbox{B)} \quad S(E, N, V) &=& \ln (c'_N\Omega (E, N, V)) = 
\ln \left(c'_N\int dq\;dp\;\theta (E - H(q,p))\right) \\ 
&=& \ln \left(c'_N{\int_{H\leq E}d\Gamma}\right)
\end{eqnarray*}  
In these formulae we have set the Boltzmann constant 
$K_B=1$, while $c_N$ and $c'_N$ are arbitrary constants to make 
the argument of the logarithm dimensionless.

Definition $A$ includes all the microstates compatible
with the constrain $H=E$, i.e. those microstates belonging to the constant
energy surface $\Sigma_{E}=\{(q ,p ) \in \Gamma \mid H(q ,p ) = E \}$. 
Alternatively, definition $B$ 
considers as microstates all those contained inside the hypervolume 
$V_E = \{(q ,p )\in\Gamma \mid H(q ,p ) \leq E \}$ 
limited by $\Sigma_E$. Both definitions should be 
equivalent in the thermodynamic limit.

Let us recall that the phase-space $\Gamma = R^{2N}$
is given the structure of a sympletic manifold by the fundamental
symplectic two-form $\omega_2=\sum_{i=1}^{N}dq_i\wedge dp_i$. 
The Hamiltonian $H : \Gamma \to R$ generates a vector field  $I\cdot dH$
(where $I$ is the fundamental symplectic matrix) and an
associated flux that preserves the Liouville measure. 
For smooth potentials $V(\{q_i\})$ it seems reasonable to assume that 
$V_E$ is limited and that $\Sigma_E$ is smooth and connected.

Taking into account case $A$, one can easily obtain an
expression for the microcanonical temperature, that from
here on we shall denote with $T_{\mu}$, through the relation
\begin{equation}
{1\over {T_{\mu}}}=
{\partial S\over\partial E}= 
{{\derome}\over\omega}
\label{Tmu} 
\end{equation} 
Making use of a simple theorem of differential geometry one has
\begin{equation} 
{1\over{T_{\mu}}}= 
{\int_{H=E}\firugh\dekinchin\over \int_{H=E}\dekinchin}= 
\<{\firugh}\>_{\mu}
\end{equation}
The derivation and the full expression of this formula are 
reported in the Appendix.

Analogously one can obtain an explicit formula for
the specific heat per particle at constant volume
\begin{equation} 
c_A =
{1\over N}\left({\partial T\over\partial E}\right)^{-1}=
{1\over N}\left(1-{\omega^2\over{\left(\derome\right)^2}}\cdot
 {{\derrome}\over\omega} \right)^{-1}
\label{CVmu}
\end{equation}
By applying the differential geometry theorems reported in the
Appendix the following expression is obtained
\begin{eqnarray}
c_A & = &{1\over N}\left(1-{\left(\int_{H=E}\dekinchin\right)^2\over 
                   \left(\int_{H=E}\firugh\dekinchin\right)^2}\cdot
                   {\left(\int_{H=E}\psirugh\dekinchin\right)\over
                   \left(\int_{H=E}\dekinchin\right)}\right)^{-1} \nonumber \\ 
    & = &{1\over N}\left(1-{\<\psirugh\>_{\mu}\over\<\firugh\>^2_{\mu}} \right)^{-1} 
\label{CVmu1}
\end{eqnarray} 

Similarly, one can treat case $B$. 

The temperature (that from here on
we denote $T_{kin}$, for reasons that will be clear in a moment) can
be obtained by introducing a vector $\eta = (0,\ldots ,0,x_i,0\ldots ,0)$,
that has the property $\nabla\cdot\eta =1$, and using the divergence theorem:
\be
\label{Tkin}
{1\over T_{kin}}={\partial S\over \partial E}={\omega\over\Omega} 
       = {\int_{H=E}\dekinchin\over\int_{H<E}\nabla\cdot\eta dx} 
       = {\int_{H=E}\dekinchin\over\int_{H=E}\gradh\cdot\eta d\sigma} 
       = {\int_{H=E}\dekinchin\over\int_{H=E} x_i{\partial H\over
        \partial x_i}\dekinchin}
\ee
This is the usual expression deriving form the virial theorem
\begin{equation} 
T_{kin}  = 
\<x_i{\partial H\over\partial x_i}\>_{\mu} 
\end{equation}
that, for Hamiltonian (\ref{hamilton}) and for
$\eta = 1/N~(0,\ldots ,0,p_1,\ldots ,p_N)$,
simplifies to
$$T_{kin}=<{\pquadro}>_{\mu}$$ i.e. the same expression that
is obtained in the canonical ensemble. One could naively
conclude that case $B$ should correspond to the canonical
ensemble for any explicit expression of thermodynamic variables.
Clearly this is not the case, as one can easily conclude 
from the expression of the specific heat per particle
\begin{equation}
c_B =
{1\over N}\left(1-{\Omega\over\omega}\cdot {{\derome}\over\omega}\right)^{-1}
={1\over N}\left(1-\< x_i{\partial H\over x_i}\>_{\mu}\cdot\<\firugh\>_{\mu}
\right)^{-1}
\label{CVkin}       
\end{equation}       
Notice that $c_B$ depends on the
ratio between $T_{kin}$ and $T_{\mu}$, that is quite a
different expression w.r.t. the corresponding canonical observable.

\section{Numerical Analysis} 

In this Section we study the dynamical and statistical properties
of the microcanonical observables introduced in Section II.
We consider the so-called FPU $\beta$-model, whose Hamiltonian
is of the form (\ref{hamilton}) with
\begin{equation}
V(\{ q_i\}) = {1\over 2}\sum_{i=0}^N(q_{i+1} - q_i)^2 + 
             {\beta\over 4}\sum_{i=0}^N(q_{i+1} - q_i)^4
\label{FPUpot}
\end{equation}
We impose fixed boundary conditions, in such a way that the total energy
$E$ is the only conserved quantity. All numerical experiments
hereafter reported were performed with $\beta = 0.1$. (Notice that
this is not a prejudice of generality, since rescaling $\beta$
amounts to rescale $E$). The integration of the
equations of motion has been performed by a bilateral symplectic
algorithm \cite{Case} in double precision, with a time step $\Delta t = 0.005$,
that guarantees the conservation of the energy at least on the sixth
significative figure, for the considered range of energies.
The running time average for an observable $f$ and the corresponding
variance, computed over the whole time span, are defined as follows:
\be
\< f \> _t= {1\over t}\int_0^t f(p(t'),q(t'))dt'
\label{mediat}
\ee
\be
{\cal V}(f) = {\< f^2 \> - \< f \>^2\over \<f \> ^2}
\label{var}
\ee
Numerical simulations typically have been performed over 
$O(10^8)$ time steps, after a transient of $10^5$ time steps.
Since in the bilateral symplectic
algorithm each time step amounts to $2\cdot \Delta t$, the total
integration time in natural units is $O(10^6)$, that corresponds 
approximately to $O(10^5)$ minimum harmonic periods of oscillations.
We have considered different values of the
energy density $\varepsilon = E/N$ and, at fixed 
$\varepsilon$, different values of $N$.
Let us first analyze the strongly chaotic regime that,
according to the results of ref.s
\cite{CLP,ACM93,ACM95,ABCM},
should extend above $\varepsilon \sim O(1)$.
For the sake of space we report here only
the running time averages of $T_{\mu}$ and $T_{kin}$ 
for $\varepsilon = 10$ and for $N = 2^n , n = 5, \cdots , 9 $ 
(see Fig. \ref{fig1}).
Notice that both $\< T_{kin} \>_t$ and
$\< T_{\mu} \>_t$  approach an asymptotic value depending on $N$:
the former observable exhibits a fast convergence
irrespectively of $N$, while the latter is characterized by wild
fluctuations that tend to weaken for increasing values of $N$.
Averaging over many different initial conditions allows one
for reducing such wild fluctuations and to extract more reliable
estimates of the asymptotic values attained by $\< T_{\mu} \>_t$.
Both "temperatures" are found to 
converge (from below and from above, respectively) 
to the same thermodynamic limit value ($T_{\infty} \approx 11.61$) with
corrections of $O(1/N)$, as it should be expected on the basis
of the ergodic theory (see Fig. \ref{fig2}). We have also checked 
that the corresponding variances ${\cal V}(T_{kin})$ and ${\cal V}
(T_{\mu})$ scale both like $1/N$ thus showing that they are characterized 
by robust statistical properties. It is worth stressing that even for 
such a high value of $\varepsilon$ the model is far from ergodic in a 
strictly mathematical sense \cite{ACM95}.
On the other hand, both temperatures seem to provide 
equally reliable thermodynamic predictions. 
We have verified that the same qualitative behaviour is observed
for values of $\varepsilon$ ranging in the interval (1, 100). 
On this basis one can state that, at least for high values of 
$\varepsilon$, the two
definitions of temperature are thermodinamically (even if not
dynamically) equivalent, while remarking that measurements of $T_{kin}$
are much more efficient for any practical purpose.

A very different scenario is obtained for $\varepsilon\ll O(1)$,
i.e. in the weakly chaotic regime. As an example,
the running time averages of $T_{\mu}$ and $T_{kin}$ 
for $\varepsilon = 0.01$ and $N = 128 $ are shown in 
Fig. 3, for three different initial conditions.
$T_{kin}$ still fastly relaxes to an asymptotic value
independently of the initial condition. Conversely,
$T_{\mu}$ exhibits a dramatic dependence on the initial
condition, despite its fluctuations are much less wild than
in the strongly chaotic regime.
The comparison between the two temperatures is quite
illuminating. $T_{kin}$ appears as a "good" thermodynamic
observable that converges to its equilibrium value,
irrespectively of ergodicity. In the perspective of
the weak ergodic theorem by Khinchin \cite{Khin} one would expect that
at least the temporal autocorrelation function of $T_{kin}$ 
decays to zero. The behaviour of $T_{\mu}$ clearly excludes
this possibility, since its dependence on the initial
conditions shows that the trajectories over which both
quantities have been measured remain "trapped" for
extremely long integration times in different regions
of the phase space. As an illustration, in Fig.s 4 a,b 
we show the normalized
temporal autocorrelation functions $C_{\mu}(t)$ and $C_{kin}(t)$
of the dynamical observables defining the 
two temperatures, for $\varepsilon = 0.01$ and $N = 128$.
The corresponding power spectra clearly indicate
that the dynamics amounts to a quasi-periodic motion
with a few dominating harmonic components, whose frequencies
and amplitudes depend on the initial conditions.
If one looks at the explicit expressions for 
$T_{\mu}$ and $T_{kin}$  this result appears much less
mysterious: the former observable is sensitive to the
geometry of the phase space (mainly depending on $V$)
through its explicit dipendence on the $q_i$'s, while the
latter is sampled over the submanifold of the momenta
$p_i$'s that makes it insensitive to any geometrical
intricacy of the phase space ( a straightforward calculation shows that
averaging $T_{kin}$ over a quasiperiodic orbit yields the same prediction
of equilibrium statistical mechanics).

The measurement of $C_{\mu}(t)$ and $C_{kin}(t)$ in the strongly chaotic
regime provides additional elements of information,
that confirm the previous conclusions and allow one
for obtaining a quantitative characterization of
the typical relaxation times of both temperatures
to their equilibrium values. In all numerical
simulations hereafter reported $C_{\mu}(t)$ and $C_{kin}(t)$
have been averaged over $5 \cdot 10^5$ initial conditions.
Typical results are shown in
Fig.s 4 c,d for $\varepsilon = 10$ and $N = 256 $.
Both autocorrelation functions exhibit a fast exponential decay
modulated by an oscillation that is rapidly damped for
large times. At variance with $C_{kin}(t)$,
$C_{\mu}(t)$ presents also a typical "hydrodynamic tail" that
superposes to the exponential decay at large values
of $t$. This shows that, even in the strongly chaotic
regime, the two temperatures still
exhibit quite different dynamical features that
originate from their different geometrical nature,
making them sensitive or insensitive to the geometrical
structure of the phase space.

A systematic inspection in a wide range of
$\varepsilon $ in the strongly chaotic regime and for 
$N = 2^n , n = 5, \cdots , 9 $ yields the following conclusions.
The hull of the autocorrelation functions can be confidently 
fitted by the
following laws
\be
C_{kin}(t)=e^{-{t/ \tau_1}}
\label{corrkin} 
\ee
\be
C_{\mu}(t)=Ae^{-{t/ \tau_1}} + {1-A\over 1+{t/ \tau_2}}
\label{corrmc} 
\ee 
where $\tau_1$ is the same for both quantities and results
to be independent of $N$, as well as $\tau_2$, and $A$ is
just a normalization constant. The main
outcome of this analysis is that, when varying $\varepsilon $
in the strongly chaotic regime,
$\tau_1$ is found to obey the following remarkable scaling law
\be
\tau_1 \propto ( \varepsilon  - \varepsilon _c )^{-1/2} 
\label{scalaw} 
\ee
with $\varepsilon_c \approx 0.8$ (see Fig. 5). It is worth
stressing that this result allows to locate unambigously
the SST of the FPU $\beta$-model at a specific value of the
energy density that agrees with all the estimates
obtained by the methods mentioned in the Introduction.
We want to observe that also $\tau_2$ is found to
diverge at $\varepsilon_c$, altough the poor numerical accuracy
in fitting the "hydrodynamic" decay of $C_{\mu}(t)$
does not allow, in this case, to extract a clean scaling law.

The same analysis has been performed for the observables $c_A$,
$c_B$ and for the canonical specific heat per particle obtained
in \cite{LPV} as a function of the microcanonical fluctuations
of the total kinetic energy $K$
\be
c_{can} \simeq \left(2 - N {{\<K^2\> - \<K\>^2}\over{\<K\>^2}}\right)^{-1}
\label{ccan}
\ee
The running time averages of $c_A$ and $c_{can}$ exhibit 
behaviours very similar to those reported in Fig. 1 for 
$T_{\mu}$ and $T_{kin}$, respectively. For the sake of space, 
in Fig. 6 we simply report the scaling with $N$ of the 
asymptotic values attained by $c_A$ and $c_{can}$, that are found
to converge to the same thermodynamic limit value $c_{\infty}$.
We have also looked at the temporal autocorrelation function of
those observables, 
whose averages determine the expressions of the
specific heats per particle. In practice, the only new interesting
quantity, with respect to those already considered in the study
of the temperatures, is the numerator of the fraction in eq.(\ref{CVmu1}).
We have verified that its autocorrelation functions coincide with
those of $T_{\mu}$ for each value of $\varepsilon$. 
All these results indicate that these specific heats provide a 
scenario fully consistent with the temperatures analysis.

On the contrary, the running time averages of $c_B$ are
characterized by extremely wild fluctuations, that do not show any
tendency to a smooth approach to the expected thermodynamic
limit value over available time spans. This fact definitely
enforce the conclusion that relaxation mechanisms may strongly
depend on the very nature of the observable at hand.

\section{Conclusions and Perspectives}

Upon assuming ergodicity, thermodynamic variables in the microcanonical 
ensemble can be explicitly written as dynamical quantities, by exploiting 
standard methods of differential geometry. In the thermodynamic limit
they can be shown to coincide with the corresponding quantities obtained in the
canonical and gran-canonical ensembles (this can be easily checked at least
for some basic quantities like temperature, specific heat etc.).

On the other hand, many degrees of freedom hamiltonian systems 
are known not to be ergodic even at high
energies \cite{ACM95}. In particular, for sufficiently
small energy densities such models exhibit quasi-periodic
evolution along trajectories, that remain trapped inside some
submanifold of the phase space for extremely long integration
times. This peculiar dynamical feature is at the basis of the
long-standing problem of ``ergodicity threshold'' raised by
the celebrated paper of FPU \cite{FPU}~. An up to date summary
of this problem is contained in the introduction.

In this paper we have pointed our attention on the 
FPU $\beta$-model, providing an unambiguous description of the 
Strong Stochasticity Threshold. This was obtained
using the previously mentioned microcanonical
observables as probes of ergodicity. In particular,
this analysis allowed us for locating
the SST at a finite value of the energy density
$\varepsilon_c \simeq 0.8$, that results to be independent of $N$.
It is worth stressing that this value agrees with rougher 
estimates still originating from the study of ``geometrical''
observables \cite{CLP,ABCM}.

Moreover, the comparison
with the corresponding canonical observables allows us for
concluding that the crucial point about the ergodic
hypothesis concerns the very nature of the observable and not just
the features of the dynamics. Such a point of view was already
raised by Khinchin \cite{Khin}, although we have shown that it may
exist thermodynamic observables (like $T_{kin}$ and $c_{can}$)
that converge to the predictions of equilibrium statistical
mechanics even when time averages are performed over highly
non-chaotic evolutions, where the ``weak ergodic theorem'' by
Khinchin does not certainly apply.

Let us finally observe that one can obtain other different 
dynamical expression for microcanonical observables by
proper choices of the coordinates space representation. The only constraint
for such a choice is to maintain invariant the flux through
the constant energy surface. For instance, in case B, one could
choose the vector $\eta = 1/2N(q_1,\cdots,q_N,p_1,\cdots,p_N)$,
in such a way that the temperature depends also on the $q_i$ coordinates.
This arbitrariness, obviously, does not affect thermodynamic limit
properties, since all expressions for the same observables coincide, 
while, dynamically, the scenario should not change significantly 
with respect to the one described in Section III.
 
Nonetheless a detailed investigation of these different choices
can provide a deeper insight on the way  ``statistical fluctuations''
can be consistently introduced in the microcanonical ensemble.
This point goes beyond the aims of this paper and will be discussed
elsewhere. 

\acknowledgments
We are indebted with H.H. Rugh, M. Pettini, S. Lepri, L. Casetti, R. Franzosi,
A. Franchi, M. Peyrard and T. Dauxois for useful discussions and comments.
We want to thank I.S.I. in Torino for the kind hospitality during
the workshop of the EU HC$\&$M Network ERB-CHRX-CT940546 
on "Complexity and Chaos", where part of this work was performed.
\appendix
\section{}

In this appendix we report a simple theorem of differential geometry
that allows to express the derivatives of the entropy  w.r.t.
the energy E. Moreover, we give the explicit
expressions for the temperature and of the quantities defining 
the specific heat of standard Hamiltonian (\ref{hamilton}).

{\bf Theorem:}
Consider $\Sigma_E$ as a $(2N-1)$-dimensional hypersurface of $R^{2N}$ 
parametrized by $H(x) = E$, where $H(x ):R^{2N}\rightarrow R$
is the Hamiltonian function in the $\Gamma$-space. 
Given a function $f(x ):R^{2N}\rightarrow R$, that is 
$C^{2N}(R^{2N})$, we let
\be\label{ap1}
\omega (E)={\int_{H=E}f(x)d\sigma}                         
\ee         
where $d\sigma$ denotes the $(2N-1)$-dimensional measure.
Suppose that it exists a constant $c>0$ such that 
$\|\nabla H\|=\<\nabla H,\nabla H\>^{1/2}\geq c$.
Then the following formula holds for the $n$-th derivative
of $\omega (E)$:
\be\label{ap2}
{d^n\omega \over dE^n}={\int_{H=E}A^nf(x)d\sigma}               
\ee
where $A^n$ denotes the $n$-th iterate of the operator $A$, defined by
\be\label{ap3}
Af(x)=\nabla\cdot\left(f(x){\nabla H\over \|\nabla H\|}\right )
{1 \over \|\nabla H\|}
\ee

{\bf Proof:} It clearly sufficies to prove the result for $n=1$.
We assume, without loss of generality, that $\nabla H$ points
towards the inside of the hypersurfaces parametrized by $H(x )=E$.
Consider the difference quotient for $\omega (E)$
\be
{\omega (E+h) - \omega(E)\over h}={1\over h}\left[{\int_{E+h}
f(x)d\sigma} - {\int_{E}f(x)d\sigma}\right]
\ee                  
Exploiting the identity 
$$1 = {\nabla H\cdot \nabla H\over \|\nabla H\|^2}$$
and, remembering that $n (x )={\nabla H\over \|\nabla H\|}$ 
is the unit inner normal to $\Sigma_E$ at point $x$, 
one can write
\begin{eqnarray}
{\omega (E+h) - \omega(E)\over h} & = & {1\over h}\left[\int_{E+h}f(x)
             \gradh n d\sigma - \int_E f(x)\gradh n d\sigma\right] \nonumber \\
     & = & {1\over h}\left[\int_{E+h}f(x)\gradh n_e d\sigma + \int_E f(x)
             \gradh n_e d\sigma\right] 
\end{eqnarray}
where $n_e$ is the unit normal pointing toward the exteriors of
the annular region \par\noindent
$\{x\in R^{2N}: E<H(x)<E+h\}$.
We may now applay the divergence theorem to obtain
\be
{\omega (E+h) - \omega(E)\over h}={1\over h}\int_E^{E+h}\nabla\cdot
               \left(f(x)\gradh\right)dx
\ee
Finally, using the co-area formula (see \cite{Khin}), we have 
\be
{\omega (E+h) - \omega(E)\over h}={1\over h}{\int_E^{E+h}dE'\int_{H=E'}
               \nabla\cdot\left(f(x)\gradh\right){\dekinchin}}
\ee
and, taking the limit for  $h\rightarrow 0$, we conclude that
\be
\omega'(E)=\int_{H=E}\nabla\cdot\left(f(x)\gradh\right){1\over\|\nabla H\|}
            d\sigma ~~~~~~~~~~~~~~~~~~~~q.e.d.
\ee 
 
\vskip .2 cm
If we consider the theorem with 
$$f(x)={1\over \|\nabla H\|}$$ 
and the cases $n=1,2$ we obtain the expressions that enter in 
the formula for temperature and specific heat given in the
previous Sections. In particular:
\be
{\derome}=\int_{H=E}\firugh\dekinchin
\ee 
\be
{\derrome}={\int_{H=E}\psirugh\dekinchin}
\ee
In the case of standard Hamiltonian (\ref{hamilton}),
for which one has $\partial H/ \partial p_i = \partial V/ \partial q_i = -F_i$  
and $\partial H/ \partial p_i=p_i$, 
the explicit dynamical expressions are:
\begin{eqnarray} 
\firugh &=& {\bigtriangleup H\over\|\nabla H\|^2}+
  \nabla\left({1\over\|\nabla H\|^2}\right)\cdot\nabla H \nonumber \\
        &=& {N-\sum_{i=1}^N{\partial F_i\over\partial q_i}\over
      \sum_{i=1}^N  p_i^2 +F_i^2} 
       +  2\cdot {\sum_{k=1}^N\sum_{i=1}^N F_k F_i 
      {\partial F_i\over\partial q_k} - \sum_{k=1}^N p_k^2\over
       (\sum_{i=1}^N p_i^2 +F_i^2)^2} 
\end{eqnarray}
\be  
\psirugh={(\bigtriangleup H)^2\over\|\nabla H\|^4}
          + \nabla\left({\bigtriangleup H\over \|\nabla H\| ^4}
          \right)\cdot\nabla H +\bigtriangleup\left({1\over\|\nabla H
          \| ^2}\right)  
\ee  
$${(\bigtriangleup H)^2\over\|\nabla H\|^4}=
   {\left(N-\sum_{i=1}^N{\partial F_i\over\partial q_i}\right)^2\over
          \left(\sum_{i=1}^N p_i^2 +F_i^2\right)^2}$$
\begin{eqnarray*}
\nabla\left({\bigtriangleup H\over\|\nabla H\|^4}
          \right)\nabla H 
      & = & {\left(\sum_{k=1}^N\sum_{i=1}^N F_k {\partial ^2 
          F_i\over\partial q_k\partial q_i}\right)\over \left(\sum_{i=1}^N 
          p_i^2 +F_i^2\right)^2} \\
      & + & 4\cdot {\left(N-\sum_{i=1}^N{\partial F_i\over\partial q_i}\right)                                     
          \left(\sum_{k=1}^N\sum_{i=1}^N F_k F_i {\partial F_i\over\partial 
          q_k}-\sum_{k=1}^N p_k^2\right)\over \left(\sum_{i=1}^N p_i^2 +F_i^2
          \right)^3}
\end{eqnarray*}
\begin{eqnarray*}
\bigtriangleup\left({1\over\|\nabla H\| ^2}\right) 
    & = & -2\cdot {\left(N + \sum_{k=1}^N\sum_{i=1}^N ({\partial F_i \over
          \partial q_k})^2 +(F_i {\partial ^2 F_i \over\partial q_k^2})\right)
          \over\left(\sum_{i=1}^N p_i^2 +F_i^2\right)^2} \\
    & + & 8\cdot {\left(\sum_{k=1}^N p_k^2 +\sum_{k=1}^{N}\left(\sum_{i=1}^N 
          F_i {\partial F_i\over\partial q_k}\right)^2\right)\over             
          \left(\sum_{i=1}^N p_i^2 +F_i^2\right)^3}
\end{eqnarray*} 
\newpage
{\bf Figure Captions}
\begin{figure}[h]
\caption{Running time averages for the temperatures $T_{kin}$ (solid
lines) and $T_{\mu}$ (dot-dashed lines) for $\varepsilon = 10$ and
$N=2^n, \quad n=5,\cdots ,9$. Increasing $N$, $T_{kin}$ ($T_{\mu}$)
better and better approximates the equilibrium expectation
from below (above).}
\label{fig1}
\end{figure}
\begin{figure}[h]
\caption{The inverse reduced temperature IRT=$|T(N)-T_{\infty}|^{-1}$
versus $N$ for $T_{\mu}(N)$ (circles) and $T_{kin}(N)$ (triangles).
Solid lines are best fits.}
\label{fig2}
\end{figure}
\begin{figure}[h]
\caption{Running time averages for the temperatures $T_{kin}$ (solid
lines) and $T_{\mu}$ (dot-dashed lines) for $\varepsilon = 10^{-2} $,
$N = 128$ and for three different initial conditions.} 
\label{fig3}
\end{figure}
\begin{figure}[h]
\caption{The temporal autocorrelation functions of $T_{kin}$ and 
$T_{\mu}$ for $N = 256$ and $\varepsilon = 10^{-2}$ ( a and b, respectively)
and for $\varepsilon = 10 $ (c and d , respectively) }
\label{fig4}
\end{figure}
\begin{figure}[h]
\caption{Log-log plot of $\tau_1^{-2}$ versus $(\varepsilon - 
\varepsilon_c)$. The solid line is the best fit obtained
assuming the otimal exstimate $\varepsilon_c = 0.79$~.}
\label{fig5}
\end{figure}
\begin{figure}[h]
\caption{The inverse reduced specific heat
IRC = $|c(N)-c_{\infty}|^{-1}$
versus $N$ for $c_A(N)$ (circles) and $c_{can}(N)$ (triangles).
Solid lines are best fits.}
\label{fig6}
\end{figure}

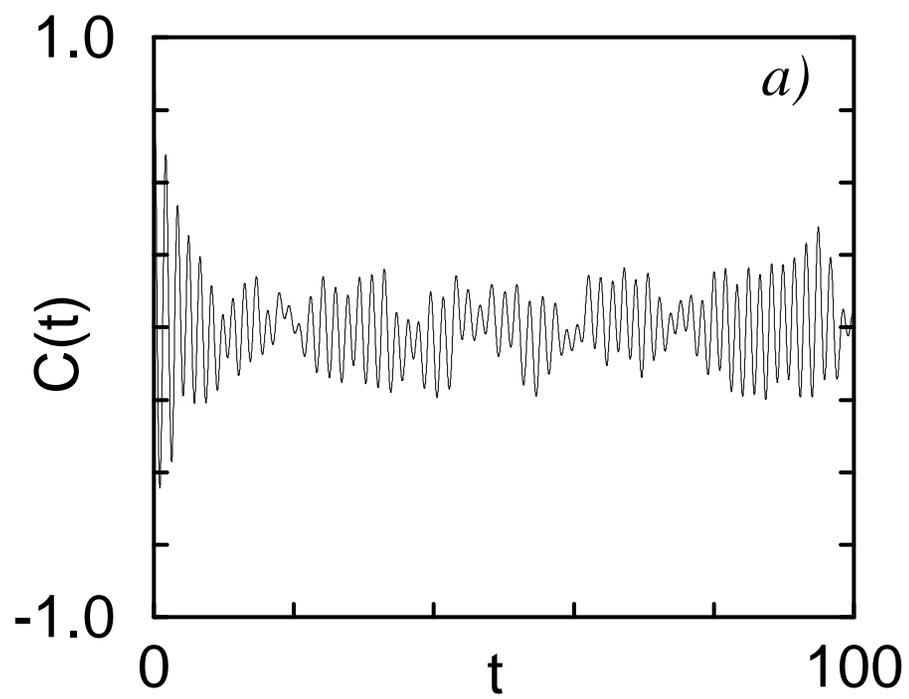
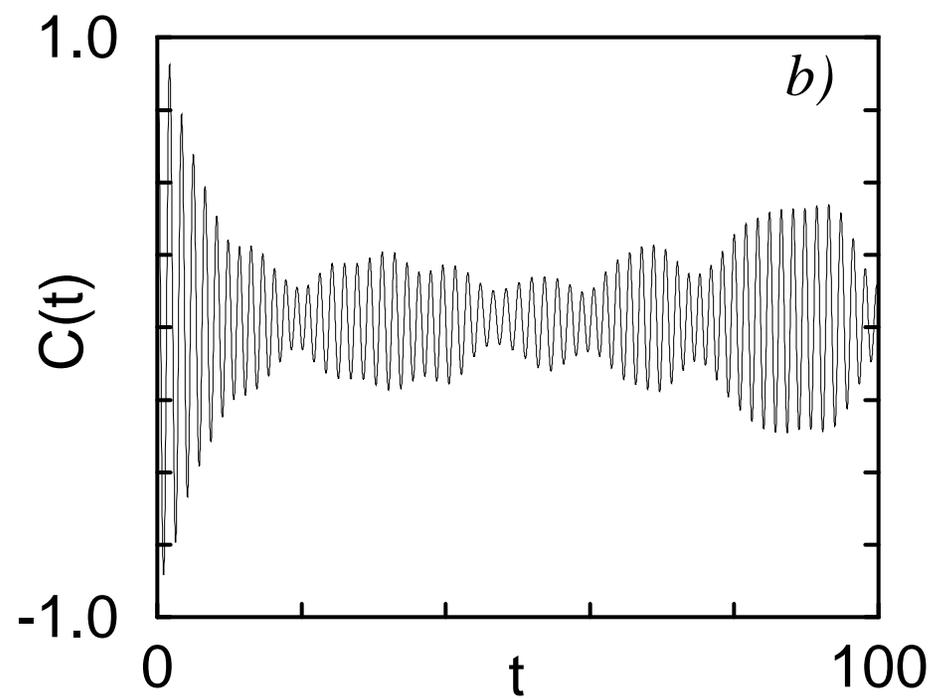
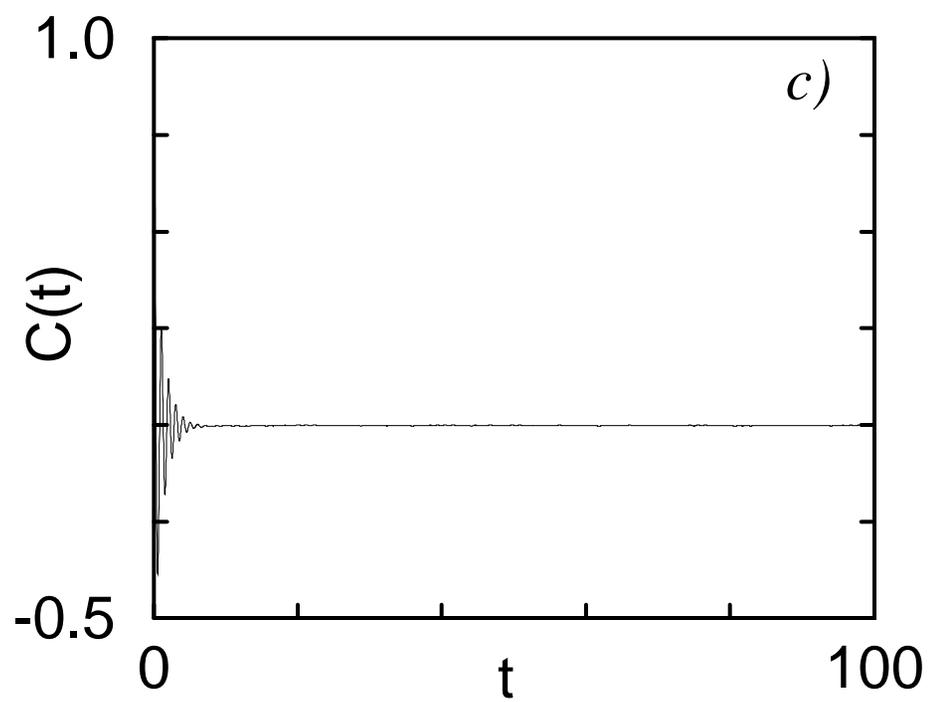
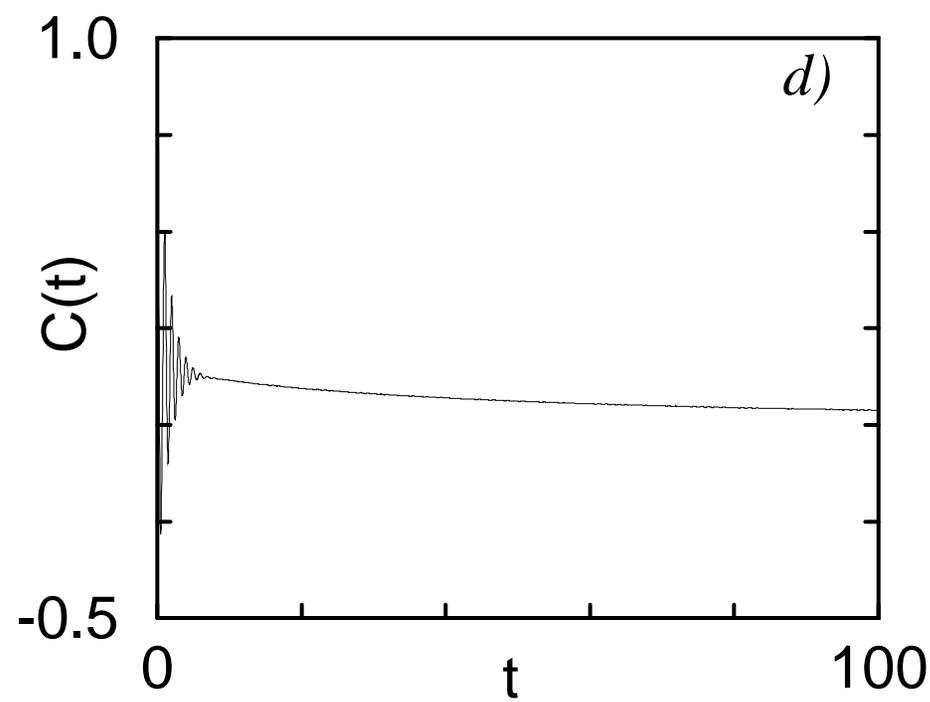
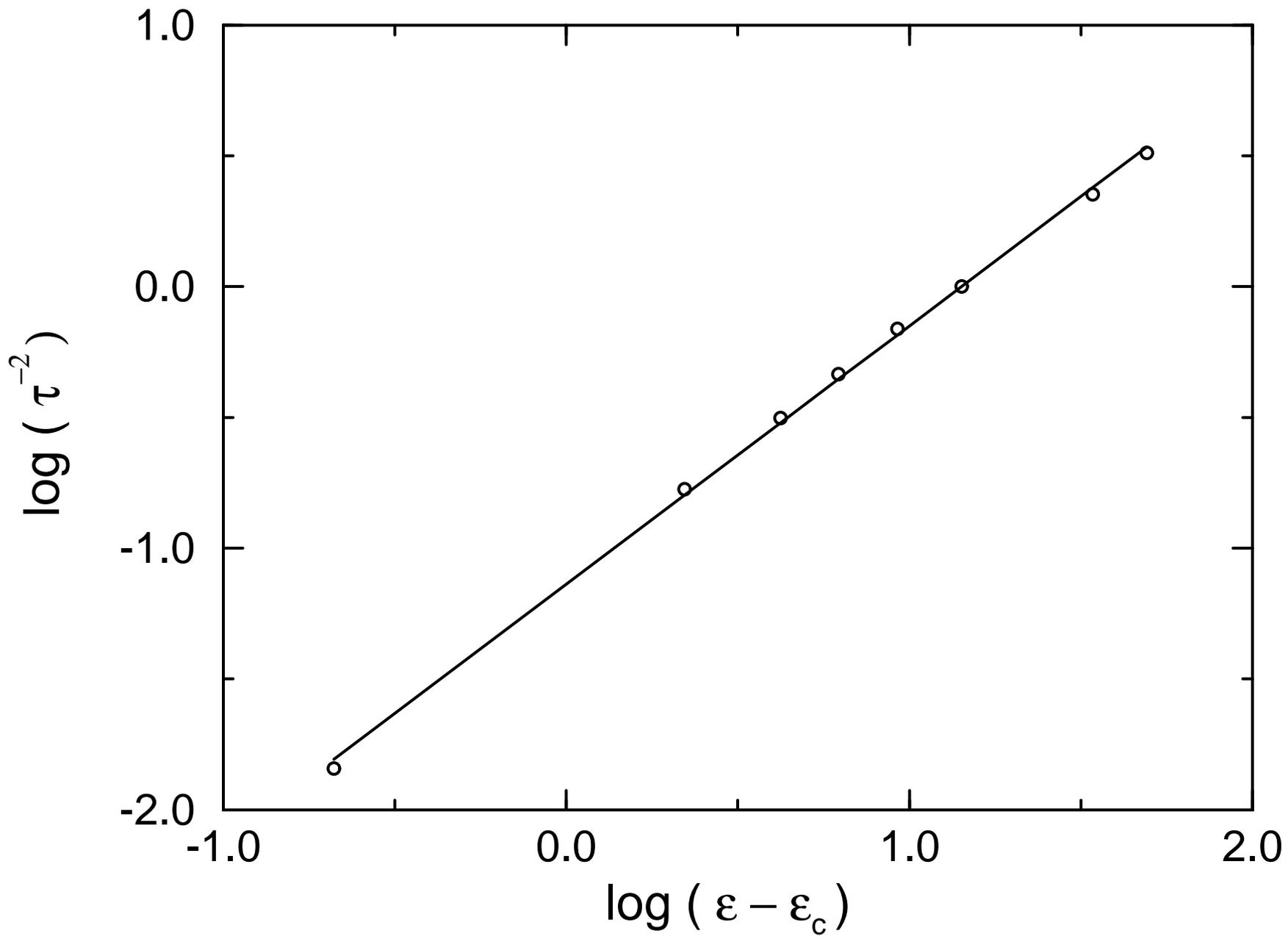
\end{document}